\documentstyle[procl,twoside]{article}
\input{psfig.sty}

\begin{document}

\setcounter{topnumber}{4}
\renewcommand\topfraction{1}

\title{Energetic particle acceleration in shear layers}

\author{Micha{\l} Ostrowski}
\address{Obserwatorium Astronomiczne, Uniwersytet Jagiello\'nski,\\ ul.
Orla 171, 30--244 Krak\'ow, Poland 
}

\maketitle\abstract{
A plasma velocity shear layer and/or a tangential flow discontinuity
provide conditions allowing for energetic particle acceleration. We
review such acceleration processes acting both in non-relativistic and
in relativistic flows. In heliospheric conditions shear layers can
provide particles with energies compatible with the observed values
(from several keV up to MeV), while in relativistic extragalactic jets
proton energies even in excess of $10^{19}$ eV can be obtained.
Application of the discussed theory to particular astrophysical objects
is severely limited by inadequate knowledge of local physical
conditions.}

\medskip
\section{Introduction}
\medskip

The first order Fermi acceleration in shock waves and the second order
acceleration in turbulent MHD media are widely considered as main
sources of cosmic ray particles in astrophysical conditions. In the
present paper we consider an alternative mechanism involving particle
acceleration at velocity shear layers formed in non-uniform plasma flows,
e.g. in a magnetosheath enveloping the Earth magnetosphere or at the
interface between the relativistic jet and its ambient medium. Till
now the considered complicated physical phenomenon was only occasionally
discussed in the literature. The process was introduced into
consideration by Berezhko and collaborators in a~series of papers in
early eighties (cf. Berezhko 1981, 1982a,b; Berezhko \& Krymsky 1983;
Bezrodnykh {\it et al.} 1984a,b, 1987; summarised in a review by Berezhko
1990). Much later an independent discussion of such processes acting in
non-relativistic shear layers was presented by Earl {\it et al.} (1988),
Jokipii {\it et al.} (1989) and Jokipii \& Morfill (1990), and for
relativistic tangential flow discontinuities by Ostrowski (1990; cf.
also Berezhko 1990). A discussion of possible cosmic ray acceleration in
mildly relativistic jets up to ultra-high energies and consequences of
acting such acceleration processes in ultra-relativistic
(`mili-arc-second') jets were considered in recent papers by Ostrowski
(1998a,b; 1999). Below, we will shortly discuss the main results
obtained by the above authors.

\section{Particle acceleration in a shear layer}

A high energy particle scattered after crossing a shear flow layer can
gain or loose energy. It is due to a respective velocity difference of
the final scattering centre rest frame with respect to the particle
starting point,
        \begin{equation}
\Delta \vec{U} = {d\vec{U} \over dx} \Delta x,
        \end{equation}
where we consider a 1-D situation with the flow velocity $\vec{U}$
directed along the $z$-axis, and the velocity gradient along the
$x$-axis of the reference frame. In absence of magnetic field $\Delta x
= v_x \Delta t$ is a free path along the $x$-axis ($\vec{v} = [v_x, v_y,
v_z]$ is the particle velocity). Let us assume for a while the
scattering centres to be static with respect to the local plasma rest
frame. Then, in the scattering centre rest frame the particle momentum
changes with respect to the one in the starting point plasma rest frame at
        \begin{equation}
\Delta p = {\Delta \vec{U} \cdot \vec{p} \over v }.
        \end{equation}
For a mean $\Delta U \ll v$ the full process can be described as
the momentum diffusion with the diffusion coefficient
        \begin{equation}
D = {1 \over 2} \left< {(\Delta p)^2 \over \Delta t} \right> = {p^2
\over 15} \left( {\partial U \over \partial x} \right)^2 \tau  ,
        \end{equation}
where the second equality comes from averaging over an isotropic
particle distribution, $\tau \equiv \left<\Delta t\right>$ is the mean scattering
time and the term $(\partial U/\partial x)^2$ is the shear scalar in the
considered simple flow pattern.

In the presence of magnetic field the mean particle shift in the $x$
direction can be much smaller than $v \, \Delta t$. Then the introduced
$\tau$ parameter equals the ratio of the particle mean free path (shift)
along the $x$-axis, $\lambda_x$, to the respective mean particle
velocity $\left< v_x\right >$, $\tau = \lambda_x /
\left<v_x\right>$. More exactly the
particle energy change and the $\tau$ parameter in Eq.~3 has to be
derived by averaging over actual particle trajectories.

\pagestyle{myheadings}
\markboth{Ostrowski}
{Energetic particle acceleration in shear layers}

If the parameter $\tau$ scales with the particle momentum as $\tau
\propto p^\eta$, then the acceleration process acting within the shear
layer produces the high energy asymptotic phase-space distribution
(Berezhko 1982)
        \begin{equation}
f(p) \propto p^{-(3+\eta)}  .
        \end{equation}
One should note that the considered acceleration process in a~shear
layer plays a~substantial role if the mean plasma velocity difference at
successive scatterings (Eq.~1) is larger than the turbulent velocities
leading to the ordinary second-order Fermi acceleration (cf. Eq.~10
below).

\section{Cosmic ray viscous acceleration in the Heliosphere}

Because of insufficient information about the turbulent shear flow
patterns within the astrophysical shear layers it is not possible to
give a firm evaluation of the viscous acceleration rates. However
numerous measurements show enhancements of energetic particle
populations in the Heliosphere, where the Solar Wind forms shear flows.
In these cases one can estimate the highest energies for accelerated
particles by the viscous mechanism and compare its to the measured ones.
Within the Heliosphere such estimates were provided for the observed
shear flow sites (cf. Berezhko 1990, Jokipii \& Morfill 1990), including
the Earth and the Jupiter sheared magnetosheath, interplanetary magnetic
field sector boundaries, boundaries of the high speed Solar Wind
streams. In general the evaluated energies are within the observed
ranges.

\section{Particle acceleration at relativistic shear layers}

The relativistic shear layers occur in a number of objects in space,
including galactic and extragalactic relativistic jets and accretion
discs near black holes. Below we consider the jet side boundary layer as
an example of the relativistic shear flow.

For particles with sufficiently high energies the transition layer
between the jet and the ambient medium can be approximated as a surface
of a discontinuous velocity change, a tangential discontinuity (`td'). It
becomes an efficient cosmic ray acceleration site provided the
considered velocity difference $U$ is relativistic and the sufficient
amount of turbulence is present in the medium. The situation with highly
relativistic jet ($\Gamma \equiv (1-U^2)^{-1/2} \gg 1$) was not
quantitatively discussed till now and, thus, our present discussion is
mostly based on the results derived for mildly relativistic flows by
Ostrowski (1990, 1998a).

\subsection{Energy gains}

Any high energy particle crossing the jet boundary changes its energy,
$E$, according to the respective Lorentz transformation. It can gain or
loose energy. In the case of uniform magnetic field the successive
transformation at the next boundary crossing changes the particle energy
back to its original value. However, in the presence of perturbations
there is a positive mean energy change:
        \begin{equation}
\left<\Delta E\right>= \eta_{\rm E} \, (\Gamma-1) \, E  .
        \end{equation}
The numerical factor $\eta_{\rm E}$ increases with the growing magnetic
field perturbations and slowly decreases for increasing $\Gamma$. For
mildly relativistic flows, in the strong scattering limit particle
simulations give values of $\eta_{\rm E}$ as substantial fractions of
unity (Ostrowski 1990). For large $\Gamma$ we assume the following
scaling
        \begin{equation}
\eta_{\rm E} = \eta_{\rm 0} {2 \over \Gamma}  ,
        \end{equation}
where $\eta_{\rm 0} = \eta (\Gamma = 2)$. In general $\eta_{\rm 0}$
depends also on particle energy. During the acceleration process,
particle scattering is accompanied with the jet's momentum transfer into
the medium surrounding it. On average, a single particle with the
momentum $p$ transports across the jet's boundary the following amount
of momentum:
        \begin{equation}
\left<\Delta p\right> =\left <\Delta p_{\rm z}\right> =  \eta_p \,(\Gamma-1) \, U
\, p  ,
        \end{equation}
where the $z$-axis of the reference frame is chosen along the flow
velocity. The numerical factor $\eta_p \approx \eta_{\rm E}$ and there
acts a drag force {\it per unit surface} of the jet boundary and the
opposite force at the medium along the jet, of the magnitude order of
the accelerated particles' energy density. Independent of the exact
value of $\eta_{\rm E}$, the acceleration process can proceed very fast
due to the fact that average particle is not able to diffuse -- between
the successive energizations -- far from the accelerating interface. One
should remember that in the case of shear layer or tangential
discontinuity acceleration and, contrary to the shock waves, there is no
particle advection off the `accelerating layer'. Of course, particles
are carried along the jet with the mean velocity of order $U/2$ and, for
efficient acceleration, the distance travelled this way must be shorter
than the jet breaking length.

The simulations (Ostrowski 1990) show that the discussed acceleration
process can be quite rapid, with the time scale given in the ambient
medium rest frame as
        \begin{equation}
 \tau_{\rm td} = \alpha \, {r_{\rm g} \over c} ,
        \end{equation}
where $r_{\rm g}$ is a characteristic value of the particle gyroradius.
For efficient scattering the numerical factor $\alpha$ can be as small
as $\sim 10$ (Ostrowski 1990). One may note that the applied diffusion
model involves particles with infinite diffusive trajectories between
the successive interactions with the discontinuity. However, quite flat
spectra, nearly coincident with the stationary spectrum (cf. Fig.~1),
are generated in short time scales given by Eq.~8 and these
distributions are considered in the present discussion. For the mean
magnetic field $B_{\rm g}$ given in the Gauss units and the particle
(proton) energy $E_{\rm EeV}$ given in EeV ($1$ EeV $\equiv 10^{18}$ eV)
the time scale (8) reads as
        \begin{equation}
\tau_{\rm td} \sim 10^5 \, \alpha \, E_{\rm EeV} \, B_{\rm G}^{-1}
\quad [{\rm s}].
        \end{equation}
For low energy cosmic ray particles the velocity transition zone at the
boundary is a~finite-width turbulent shear layer. We do not know of any
attempt in the literature to describe the internal structure of such
layer on the microscopic level (cf. Aloy {\it et al.} 1999,
Henriksen, at this meeting). Therefore,
we limit the discussion of the acceleration process within such a layer
to quantitative considerations only. From rather weak radiation and the
observed effective collimation of jets in the powerful FR~II radio
sources one can conclude, that interaction of a presumably relativistic
jet with the ambient medium is relatively weak. Thus the turbulent
boundary layer must be relatively thin, with thickness denoted with $D$.
Within it two acceleration processes energise low energy -- the ones
with the mean radial free path $\lambda \ll D$ -- particles. The first
one, discussed in section 2 above, is connected with the velocity shear
and is called `cosmic ray viscosity'. The second one is the ordinary Fermi
process in the turbulent medium. The acceleration time scales can not be
evaluated with accuracy for these processes, but -- for particles
residing within the considered layer -- we can give an acceleration time
scale estimate
        \begin{equation}
        \pagebreak[4]
\tau_{\rm II} = {2 \pi r_{\rm g} \over c} {c^2 \over V^2 + \left( U
{\lambda \over D} \right)^2 }  ,
        \end{equation}

        \begin{figure}[t]
\centerline{
\psfig{file=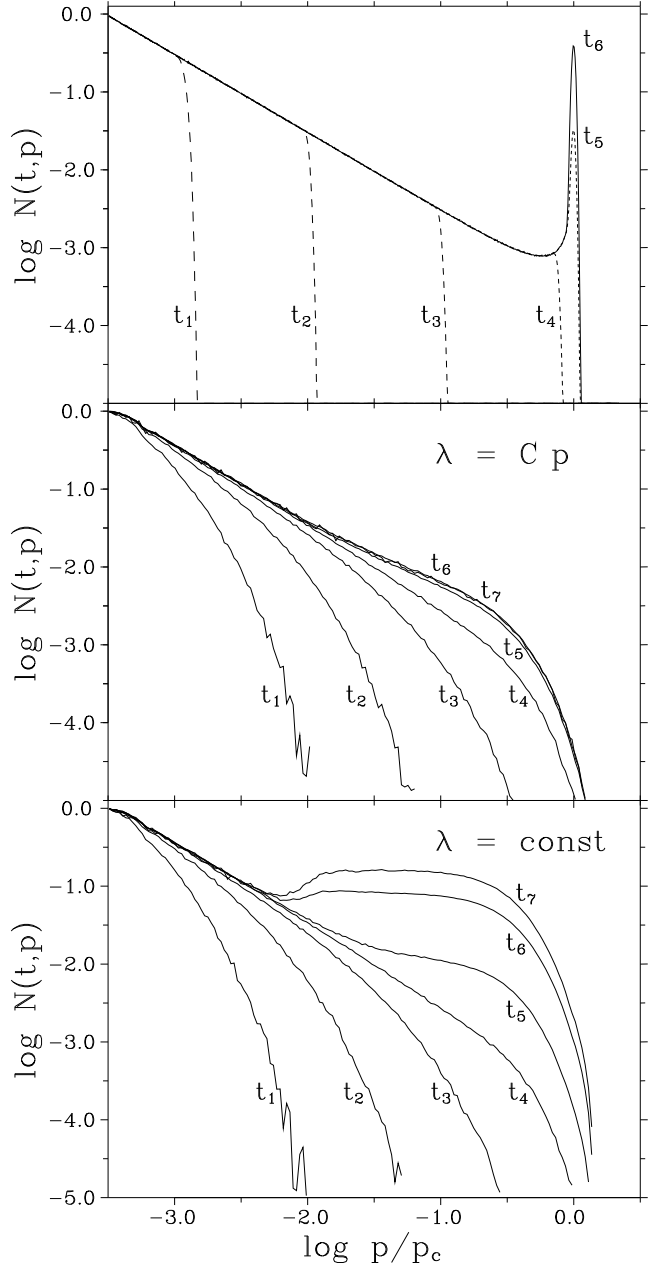,width=8.5cm,angle=0}
}
\caption{Spectra $N(t, p, x = 0 ) \equiv d\, n / d\, \log\, p$  of
accelerated particles at the jet boundary in a sequence of times $t_1 <
t_2 < \ldots < t_7$ (each $t_i = 10 t_{i-1}$), for continues particle
injection at small momentum $p_{\rm 0} \ll p_{\rm c}$. In the upper plot
the spectra generated in the turbulent shear layer are given, while, in
the two lower ones, the tangential discontinuity generated spectra are
presented: the results for the particle mean free path $\lambda \propto
p$ are given in the upper panel, and for $\lambda = {\rm const}$ in the
lower one.\hfill\mbox{}}
        \end{figure}

\clearpage

\noindent
where $V$ is the turbulence velocity. One expects that the first term
in the denominator can dominate at low particle energies, while the
second one for larger
energies, with $\tau_{\rm II}$ approaching the value given in Eq.~8
for $\lambda \sim D$. If the second-order Fermi acceleration dominates,
$\lambda < D (V/U)$, the time scale (10) reads as
$\tau_{\rm II} \sim 10^8 \, E_{\rm TeV} \, B_{\rm G}^{-1} V_3^{-2}$
$[s]$, where $V_3$ is the turbulence velocity in units of $3000$ km/s.
Depending on the choice of parameters this scale can be comparable or
longer than the expansion and internal evolution scales for relativistic
jets. In order to efficiently create high energy particles for the
further acceleration by the viscous process and the tangential
discontinuity acceleration one have to assume that the turbulent layer
includes high velocity turbulence, with $V_3$ reaching values
substantially larger than $1$, or other high energy particles sources
are present. For the following
discussion we will assume that such effective pre-acceleration takes
place, but the validity of this assumption can be estimated only {\it a
posteriori} from comparison of our conclusions with the observational
data.

\subsection{Energy losses}

To estimate the upper energy limit for accelerated particles, at first
one should compare the time scale for energy losses due to radiation and
inelastic collisions to the acceleration time scale. The discussion of
possible loss processes is presented by Rachen \& Biermann (1993). The
derived loss time scale for utra-high energy protons can be written in
the form
        \begin{equation}
T_{\rm loss} \simeq 5\cdot 10^9~B_{\rm g}^{-2} \, (1+Xa)^{-1} \,
E_{\rm EeV}^{-1} \quad [{\rm s}]  ,
        \end{equation}
where $a$ is the ratio of the energy density of the ambient photon field
relative to that of the magnetic field and $X$ is a quantity for the
relative strength of {\it p}$\, \gamma$ interactions compared to
synchrotron radiation. For cosmic ray protons the acceleration dominates
over the losses (Eqs.~9, 11) up to the maximum energy $E_{\rm EeV}
\approx 2 \cdot 10^2 \alpha^{-1} \left[ B_{\rm G} (1+Xa)
\right]^{-1/2}$.

\subsection{Spectra of accelerated particles}

The acceleration process acting at the tangential discontinuity of the
velocity field leads to the flat energy spectrum and the spatial
distribution expected to increase their extension with particle energy.
Below, for illustration, we propose two simple acceleration and
diffusion models describing these features.

\subsubsection{A turbulent shear layer}

At first we consider `low energy' particles wandering in an extended
turbulent shear layer, with the particle mean free path $\lambda \propto
p$. With the assumed conditions the mean time required for increasing
particle energy on a small constant fraction is proportional to the
energy itself, and the mean rate of particle energy gain is constant,
$\langle\dot{E}\rangle_{\rm gain} = \mbox{const.}$ Let us take a
simple expression for the
synchrotron energy loss, $\langle\dot{E}\rangle_{\rm loss} \propto p^2$, to
represent any real process acting near the discontinuity. One may note
that the jet radius and the escape boundary distance provide energy
scales to the process. Another scale for particle momentum, $p_{\rm c}$,
is provided as the one for equal losses and gains, $\langle\dot{E}\rangle_{\rm gain}
= \langle\dot{E}\rangle_{\rm loss}$. As a result, a divergence from the power-law
and a cut-off have to occur at high energies in the spectrum.

\tolerance500
We use a simple Monte Carlo simulations to model the acceleration
process for a~continuous particle injection, uniform within the
considered layer. The diffusion coefficient $\kappa_\perp$ is taken to
be proportional to particle momentum, but independent of the spatial
position $x$. We neglected particle escape through the shear layer side
boundaries and we assumed ${\partial f \over \partial x} = 0$~. For the
escape term we simply assume a characteristic escape momentum $p_{\rm
max}$. In Fig.~1 we use $p_{\rm c}$ as the unit for particle momentum, so
it defines also a cut-off for $p_{\rm c} < p_{\rm max}$~. At small
momenta the spectrum has a power-law form -- in our model the averaged
over angles $f(t,p) \propto p^{-4}$ (cf. Eq.~4) -- with a cut-off
momentum growing with time. However, in long time scales, when particles
reach momenta close to $p_{\rm c}$, losses lead to spectrum flattening
and pilling up particles at $p$ close to $p_{\rm c}$. Then, a low energy
part of the spectrum does not change any more and only a narrow spike at
$p \approx p_{\rm c}$ grows with time. Let us also note that in the case
of efficient particle escape, i.e. when $p_{\rm max} < p_{\rm c}$, the
resulting spectrum would be similar to the presented by Ostrowski (1998a)
short time spectrum with\break a cut-off at $\approx p_{\rm max}$.

\tolerance300

\subsubsection{Tangential discontinuity acceleration}

An illustration of the acceleration process at the tangential
discontinuity have to take into account a spatially discrete nature of
the acceleration process. Here, particles are assumed to wander subject
to radiative losses outside the discontinuity, with the mean free path
$\propto p$ and the loss rate $\propto p^2$. At each crossing of the
discontinuity a particle is assumed to gain a constant fraction $\Delta$
of momentum (cf. Eqs.~5, 6), $p^\prime = (1+\Delta) \,  p $~, and, due to
losses, during each free time $\Delta t$ its momentum decreases from
$p_{\rm in}$ to $p$ according to
        \begin{equation}
{1 \over p} - {1 \over p_{\rm in}} = {\rm const} \cdot \Delta t.
        \end{equation}
The time dependent energy spectra obtained within this model are
presented in lower panel in Fig.~1, where we choose units in a way to put
the constant in Eq.~12 equal to one and the particle mean free paths
are equal
in two considered models at $p = p_{\rm c}$. Comparison of the results
in two models allows to evaluate the modification of the acceleration
process by changing the momentum dependence of the particle diffusion
coefficient. For slowly varying diffusion coefficient (the `$\lambda =
{\rm const}$' model) high energy particles which diffuse far away off the
discontinuity and loose there much of their energy still have a chance
to diffuse back to be accelerated at the discontinuity. In the model
with $\kappa$ quickly growing with particle energy (the `$\lambda =
C \cdot p$' model) such distant particles will decrease their mobility
in a degree sufficient to break, or at least to limit their further
acceleration. One should note that in both models the spectrum
inclination at low energies is the same (here the particle density $n(p)
\propto p^{-2}$).

\section{Final remarks}
\vspace{-4pt}

Shear layers occurring in astrophysical plasma flows are able to
accelerate cosmic ray particles in the so called {\em viscous
acceleration process}. Depending on conditions the process can be
described as the particle momentum diffusion or the tangential
discontinuity acceleration. The last one can be very efficient in
relativistic flows. In particular, in jets in active galactic nuclei the
cosmic ray protons can reach energies in excess of $10^{18}$ eV.

The generated cosmic ray populations can influence the shear layer flow
through viscous and/or dynamical forces (cf. Arav \& Begelman 1992,
Ostrowski 1999). In relativistic jets the so called {\it cosmic ray
cocoon} can be formed leading to a number of observational effects. The
essential problem with application and verification of the presented
theory is insufficient information about the local parameters of the
considered shear layers. One may note several recent observational
papers showing effects which could be ascribed to, or are at least
compatible with the acceleration process acting at jet boundary layer
(Attridge {\it et al.} 1999, Scarpa {\it et al.} 1999, Perlman
{\it et al.} 1999).

\vspace{-4pt}
\section*{Acknowledgments}
\vspace{-4pt}

I\, acknowledge\, support\, from\, the\, {\it Komitet\,
Bada\'n\, Naukowych\/}\, within\,
project\break 2~P03D~002~17 and 2~P03B~112~17.

\vspace{-4pt}

\references
\vspace{-4pt}

Aloy M.A., Ib\'a{\~{n}}ez J. M$^a$, Marti J.M$^a$, Gomez
J.L., M\"uller E. (1999) {\em Astrophys.~J. Lett.} (accepted).

Attridge J.M., Roberts D.H., Wardle J.F.C. (1999) {\em Astrophys.~J.
             Lett.} {\bf 518}, 87.

Arav N., Begelman M.C. (1992) {\em Astrophys.~J.} {\bf 401}, 125.

Berezhko E.G. (1981) {Pisma v ZhETF} {\bf 33}, 416.

Berezhko E.G. (1982) {\em Pisma v Astr. Zh.} {\bf 8}, 747.

Berezhko E.G. (1982) {\em Geomagnetizm i Aeronomia\/} {\bf 22}, 321.

Berezhko E.G. (1990) Preprint {\it Frictional Acceleration of
             Cosmic Rays}, The Yakut Scientific Centre, Yakutsk.

Berezhko E.G., Krymsky G.F. (1983) {\em Izvestiya AN SSSR,
Seriya Fiz.} {\bf 47}, 1700.

Bezrodnykh I.P., Berezhko E.G., Plotnikov I.Ya., {\it et al.} (1984a)
         {\em Izviestiya AN SSSR, Seriya Fiz.} {\bf 48}, 2164.

Bezrodnykh I.P., Berezhko E.G., Plotnikov I.Ya., {\it et al.} (1984b)
         {\em Geomagnetizm i Aeronomia\/} {\bf 48}, 2164.

Bezrodnykh I.P., Berezhko E.G., {\it et al.} (1987) in Proc. 20th Int. Cosmic
          Ray Conf., Moscow {\bf 5}, 453.

B\"ottcher M. (1999) {\em Astrophys.~J., Lett.} (accepted).

Earl J.A., Jokipii J.R., Morfill G.E. (1988) {\em Astrophys. J. Lett.}
          {\bf 331}, L91.

Jokipii J.R., Morfill G. (1990) {\em Astrophys. J.} {\bf 356}, 255.

Jokipii J.R., Kota J., Morfill G. (1989) {\em Astrophys. J.
Lett.} {\bf 345}, L67.

Ostrowski M. (1990) {\em Astron. Astrophys.} {\bf 238}, 435.

Ostrowski M. (1998a) {\em Astron. Astrophys.} {\bf 335}, 134.

Ostrowski M. (1998b) in {\it Frontier objects in astrophysics and
particle physics (Vulcano Workshop)}, eds. F. Giovannelli
\& G. Mannocchi.

Ostrowski M. (1999) {\em Month. Not. R.~Astron. Soc.} (in press).

Perlman E.S., Biretta J.A., Fang Z., Sparks W.B., Macchetto F.D. (1999)
          {\em Astron.~J.} (accepted).

Rachen J.P., Biermann P. (1993) {\em Astron. Astrophys.} {\bf 272}, 161.

Scarpa R., Urry C.M., Falomo R., Treves A. (1999) {\em Astrophys.~J.}
          (accepted).

\end{document}